\documentclass[reprint,aps,prd,nofootinbib, floatfix]{revtex4-1}
\usepackage[utf8]{inputenc}
\usepackage[utf8]{inputenc}
\usepackage{amsmath}
\usepackage{amsfonts}
\usepackage{amssymb}
\usepackage[left=2cm,right=2cm,top=2cm,bottom=2cm]{geometry}
\usepackage{braket}
\usepackage{dsfont}
\usepackage{hyperref}

\usepackage{graphics}
\usepackage{tikz}
\usetikzlibrary{arrows,decorations.pathmorphing,backgrounds,positioning,fit,petri,shapes}
\usepackage{lipsum}

\newcommand{\red}[1]{\textcolor{black}{#1}}
\newcommand{\henk}[1]{\textcolor{black}{#1}}

\usepackage{multirow}
\begin{document}

\title{Rainbow Cosmic Shear: Optimisation of Tomographic Bins}

\begin{abstract}
In this paper we address the problem of finding optimal cosmic shear tomographic bins. We generalise the definition of a cosmic shear tomographic bin to be a set of commonly labelled voxels in photometric colour space; rather than bins defined directly in redshift. We explore this approach by using a self-organising map to define the multi-dimensional colour space, and a we define a `label space' of connected regions on the self-organising map  using overlapping elliptical disks. This allows us to then find optimal labelling schemes by searching the label space. We use a metric that is the signal-to-noise ratio of a dark energy equation of state measurement, and in this case we find that for up to five tomographic bins the optimal colour-space labelling is an approximation of an equally-spaced binning in redshift; that is in all cases the best configuration. We also show that such a redefinition is more robust to photometric redshift outliers than a standard tomographic bin selection.
\end{abstract}

\author{Thomas D. Kitching, Peter L. Taylor}
\affiliation{Mullard Space Science Laboratory, University College London, Holmbury St. Mary, Dorking, Surrey RH5 6NT, UK}
\author{Peter Capak}
\affiliation{Department of Astronomy, California Institute of Technology, MC 249- 17, 1200 East California Blvd, Pasadena, CA 91125, USA}
\author{Daniel Masters}
\affiliation{Jet Propulsion Laboratory, California Institute of Technology, 4800 Oak Grove Drive, Pasadena, CA 91109, USA}
\author{Henk Hoekstra}
\affiliation{Leiden Observatory, Leiden University, Niels Bohrweg 2, NL-2333 CA Leiden, The Netherlands}
\date{\today}
\maketitle

\section{Introduction}

Despite the tremendous progress in the precision with which cosmological parameters have been determined, we still do not understand the physical nature of the main ingredients that make up the Universe in the currently favoured $\Lambda$CDM model. To advance observational constraints a number of techniques can be employed. Of these, weak gravitational lensing by large-scale structure, or cosmic shear, is potentially the most promising. It uses the distortions in the observed images of distant galaxies to map the  mass perturbations along the line-of-sight. These projected image distortions - changes in the size and third flattening (ellipticity) of the images - are correlated in the angular direction and as a function redshift and these correlations are typically summarised in terms of a power spectrum or a correlation function. 

Thanks to ever larger, deep imaging surveys with good image quality, the lensing signal is now routinely measured \citep[e.g.][]{hildebrandt2017kids,troxel2018des, hikage2018hsc}, and yields constraints on certain parameter combinations that match the precision of the most recent CMB results \cite{planck18}. Moreover, even larger surveys aim to constrain the dark energy equation-of-state with per cent precision \citep{EuclidRB,LSST, WFIRST}. The projected matter power spectrum is rather featureless, and the sensitivity to dark energy, and modifications to the theory of gravity, arises from the ability to measure the growth of structure by dividing the source samples by redshift. Obtaining precise spectroscopic redshifts for such large samples of distant, faint galaxies is not possible, but  the lensing kernel is broad so that photometric redshifts are adequate. However uncertainties in the photometric redshift determination of individual sources prevent a clean division of the sources, thus reducing the precision with which cosmological parameters can be determined.

In practice the division is commonly done by determining photometric redshifts using the available multi-band data and binning these estimates. The more bands, the better the mapping between the colours and the redshift. It is, however, not clear what the best labelling of sources is for the binning, because of statistically ill-defined (catastrophic) outliers and the overall statistical uncertainties in the redshift determination. Instead of this approach we explore here the possibility to bypass the photometric redshift determination step, and divide the sources based on their colour measurements directly.

In this paper we investigate how one should optimally choose colour voxel sets for cosmic shear. We do this by using the self-organising map derived from \cite{masters}, so we can work in lower dimensions, and explore this space. The methodology is presented in Sections 2 and 3, results are presented in Section 4, and we present our conclusions in Section 5. This is a preliminary study of this re-definition of cosmic shear tomography, that will be extended in future works to include intrinsic alignments, galaxy-clustering cross-correlations and combined with optimal scale-cutting weights \cite{kcut}. 

\section{Cosmic Shear Tomography}
The most widely used cosmic shear power spectrum formulation uses the Limber \cite{kitchinglimits}, flat-Universe \cite{taylorflat} and equal-time \cite{kitchingunequal} approximations, and creates bins in redshift known as `tomographic' bins. This approach was generalised in \cite{taylor} to a \textit{generalised spherical-transform} that is defined as: 
\begin{equation} \label{eq:C}
\gamma_{\ell m} \left( \eta \right) =  \sqrt{\frac{2}{\pi}} \sum _g \gamma_g \left( r_g, \boldsymbol{\theta_g } \right)W_{\ell} \left(\eta, r_g \right)  {}_2 Y_{\ell m} \left( \boldsymbol {\theta_g} \right), 
\end{equation}
where $\gamma \in \mathbb{C}$ is the shear, the sum is over all galaxies $g$ with angular coordinate ${\bf \theta_g}$ and radial coordinate $r_g$, $W_\ell(\eta_i,r_g)$ is a weight where $\eta_i$ is a general label, and $_2 Y_{\ell m}$ are the spin-weighted spherical harmonics with spin $2$. The cosmic shear power spectrum is the covariance of this quantity that is:
\begin{equation} \label{eq:c_l}
C_{\ell}\left( \eta_i, \eta_j \right) =  \frac{9 \Omega_m ^ 2 H_0 ^ 4}{16 \pi^4 c^ 4 }\frac{\left( \ell + 2 \right)!}{\left( \ell - 2 \right)!} \int \frac{\text{d} k}{ {k} ^2} G_{\ell}\left( \eta_i, k \right) G_{\ell}\left(\eta_j, k \right) ,
\end{equation}
where $c$ is the speed of light in vacuum, $\Omega_m$ is the fractional mass-energy density of matter, and $H_0$ is the value of the Hubble constant. The matrix $G$ is given by:
\begin{equation} \label{eq:G}
\begin{aligned}	
G_{\ell}\left( \eta_i , k \right) \equiv \int \text{d}z_p \text{d} z' \text{ }  &n \left(z' \right) p_i \left(z' | z_p \right) \\ & \times W_{\ell} \left(\eta_i, r  \left[z' \right]\right) U_{\ell} \left(r \left[ z' \right], k \right)  
\end{aligned}
\end{equation}
where $r[z]$ is the co-moving distance at a redshift $z$. $p_i \left( z|z_p \right)$ is the probability that a galaxy has a redshift $z$, given a photometric redshift measurement $z_p$ for bin $i$. The radial distribution of galaxies is denoted by $n(z)$. By taking the the Limber and flat-Universe approximations ~\cite{loverdelimber,taylorflat}, the matrix $U$ can be written as:
\begin{equation} \label{eq:limber}
U_{\ell} \left(r, k \right) = \frac{F_K \left( r, \nu \left( k \right) \right)}{k a \left( \nu\left( k \right)  \right)} \sqrt {\frac{\pi}{2 \left( {\ell} + 1/2 \right)}}  P ^ {1/2} \left( k, \nu\left( k \right)  \right),
\end{equation}
where $\nu\left( k \right)  \equiv \frac{{\ell}+ 1/2}{k}$. These are good approximations for small-scales $\ell> 100$ \cite{kitchinglimits}, and for flat universes consistent with current measurements of $\Omega_k$ \cite{taylorflat}. The power spectrum of matter overdensities is denoted $P(k;r)$ where the equal-time approximation has been used \cite{kitchingunequal}.  The lensing kernel, for a spatially flat cosmology is:
\begin{equation} \label{eq:kernel}
 F_K \left(r, r' \right)\equiv \frac{r-r'}{rr'}.
\end{equation} 
The shot noise power spectrum, caused by the random (unlensed) ellipticity of galaxies, is given by:
\begin{equation} \label{eq:Noise}
N_\ell\left( \eta_1, \eta_2 \right) = \frac{\sigma_e ^2}{2 \pi ^ 2} \int \text{d} z \text{ } n\left( z \right)W_\ell \left(\eta_1, r \right)W_\ell \left(\eta_2, r \right) , 
\end{equation}
where $\sigma_e ^2\simeq (0.3)^2$ \cite{brown2003shear} is the variance of the unlensed ellipticities of the observed galaxies. The shot noise amplitude scales as $1/N_g$, where $N_g$ is the number of galaxies in a given bin, see \cite{kcut}.

From this general definition the question is then which weight function to choose. By taking the weight-function, $W_\ell \left(\eta_i, r  \left[ z \right]\right) \equiv j_\ell \left(\eta_i r[z]\right)$ in equations~\eqref{eq:G} and~\eqref{eq:Noise} the equations for `3D cosmic shear' are reproduced \cite{heavens3d,heavens2006measuring,castro}. In this case the labels $\eta$ correspond to inverse-distance variables in the Bessel function. The standard `tomographic' cosmic shear spectra \cite{hushotnoise}, are reproduced by taking the weight function to be a top hat function in redshift only:
\begin{equation}
\label{tomow}
   W\left(\eta_i, z \right) \equiv
    \begin{cases}
      1 & \text{if $z \in i$  }\\
      0 & \text{if $z \notin i$,  }\\
    \end{cases} 
\end{equation}
defines the `tomographic' bin associated with redshift region $i$. Normally tomographic bin selection is done in one of two ways: equally-spaced bins in redshift, or bins that have an equal-number of galaxies per bin; we refer to theses as `equally-spaced' and `equal-number' as a shorthand throughout. Both these options are discussed in \cite{taylor}.  

In this paper we generalise the `tomographic' bin labels $\eta_i$ to be indicators of a population of galaxies with similar colours (or SEDs), \red{or with SEDs close in colour space}, rather than \red{directly} similar spectroscopic redshifts (as is the case in standard tomography). We refer to the corresponding power spectra $C_{\ell}^{\gamma \gamma} \left( \eta_i, \eta_j \right)$ as `rainbow tomography'. In this sense the colour combinations for a set of bin labels $\{\eta_i\}$ can generically result in complicated equivalent behavior as a function of redshift. This is similar to the proposal of \cite{jain}. More specifically the bin labels $\eta_i$ correspond to a set of colour voxels $V$ that have been given a common label $i$: 
\begin{equation}
\eta_i\mathrel{\widehat{=}}\bigcup V_a,\dots,V_n=\{V\}_i
\end{equation}
where the union of the voxels $a\dots n$ corresponds to bin $i$. We use the notation ${\widehat{=}}$ to mean `corresponds to'. 

For each voxel $V_{\alpha}$ it is assumed that there is a specific unique spectroscopic probability distribution $n_{\alpha}(z)$ and a corresponding photometric redshift probability distribution  $p_{{\alpha}}(z|z_p)$, derived from broadband information used in a specific experiment. We note that the sum over all the voxels results in the radial  distribution of galaxies that enters equation (\ref{eq:G})$, \sum n_{\alpha}(z)=n(z)$. \red{Within a voxel the mapping of spectroscopic redshift to photometric redshift probability distribution is a subject of photometric redshift estimations codes such as \cite{BPZ,LP}}. Therefore we define the weight function for a given label $i$ to be 
\begin{equation}
\label{weight}
W\left(\eta_i, z \right)=\frac{1}{n(z)}\sum_{\alpha \in \{V\}_i} n_{\alpha}(z), 
\end{equation}
where the sum is over all voxels in the set $\{V\}_i$. We note that the standard tomographic bin definitions (equally-spaced, and equal-number) are instances of this definition where, in reference to equation (\ref{tomow}), the voxels in $\{V\}_i$ correspond to all galaxies with $z_{\alpha} \in i$. The denominator ensures that in equation (\ref{eq:G}) this definition reduced to the weights used in the standard tomographic case (equations \ref{tomow}, \ref{eq:Noise}). 

For reference, \red{one can define the photometric redshift probability distribution for a given bin 
\begin{equation}
\label{sweight}
p_i(z,z_p)=\frac{1}{N_i}\sum_{\alpha \in \{V\}_i} p_{\alpha}(z|z_p), 
\end{equation}
where each $0<p_{\alpha}(z|z_p)<1$ is an individual probability distribution for voxel $\alpha$, and the denominator is the number of voxels $N_i$ that contribute to bin $i$ which ensures that $0<p_i(z,z_p)<1$.
}

\section{Methodology} \label{sec:methodology}
Here we briefly review the self-organising map approach of \cite{masters} and then describe how this space is explored to optimally select colour voxel combinations for tomographic bin labelling. 

\subsection{Self-Organising Map}
\label{Self-Organising Map}
% show what the equal-z and equal-n maps look like 
% define the region parameterisation
When observing galaxies in $N_B$ broadband filters the colour space (all possible wavelength differences in this set) is of dimension $N_B!/2$, that for cosmic shear surveys results in a high-dimensional space. In this space  colour combinations that have similar spectroscopic redshifts form lines or planes (see e.g. \cite{masters}). In order to more efficiently represent this space \cite{masters} applied a self-organising map that projects this high-dimensional data onto a lower dimensional (2D) manifold. The result is a map of 2D pixels that represent voxels in the higher-dimensional space, and crucially that the topology of the higher-dimensional space is retained. In \cite{masters} this method was used to map the multi-colour space and its completeness using COSMOS data, in particular to determine in which pixels additional spectroscopic information is required. 

\begin{figure}
\centering
\vspace{2mm}
\includegraphics[width=\columnwidth]{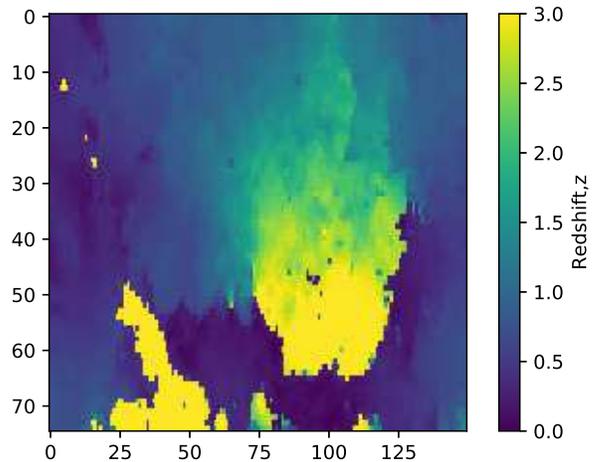}
\caption{The \cite{masters} self-organising map. The colour scale shows the mean \red{photometric} redshift per pixel. The axes are pixel labels that are arbitrarily defined.}
\label{fig:som_z}
\end{figure}

In Figure \ref{fig:som_z} we show \red{a} 2D self-organising map, coloured by the mean redshift per pixel, the axes are pixel numbers that are arbitrary labels. \red{This map is a  modified version of the one used in \cite{masters} made for C3R2 targeting, that now includes VVDS/EGS in its derivation as well as COSMOS; this will be described in the C3R2 DR2 paper, (Masters et al., in prep).}

This represents the redshift distribution for  Euclid\footnote{\url{http://euclid-ec.org}} \cite{EuclidRB}. For each pixel there is in principle a full posterior redshift distribution, see equation (\ref{sweight}), but the current version has only mean redshift estimates; this is sufficient for our proof of concept study where for each bin we assign a Gaussian $p_{{\alpha}}(z|z_p)$ with mean $z_p$ and width $\sigma(z)=0.03(1+z)$.

We note several features of the self-organising map that are salient to the discussion in this paper. Firstly neighbouring pixels represent voxels in colour space that have similar spectral energy distributions (SEDs). This means that areas of interconnected and adjacent pixels can be delineated in this space that have some physical meaning. Secondly there is some structure in this space, i.e. non-random and well defined in shape, suggesting that the topology of the projected 2D space can be approximated using simple shapes.   
\begin{figure*}
\centering
\includegraphics[width=0.65\columnwidth]{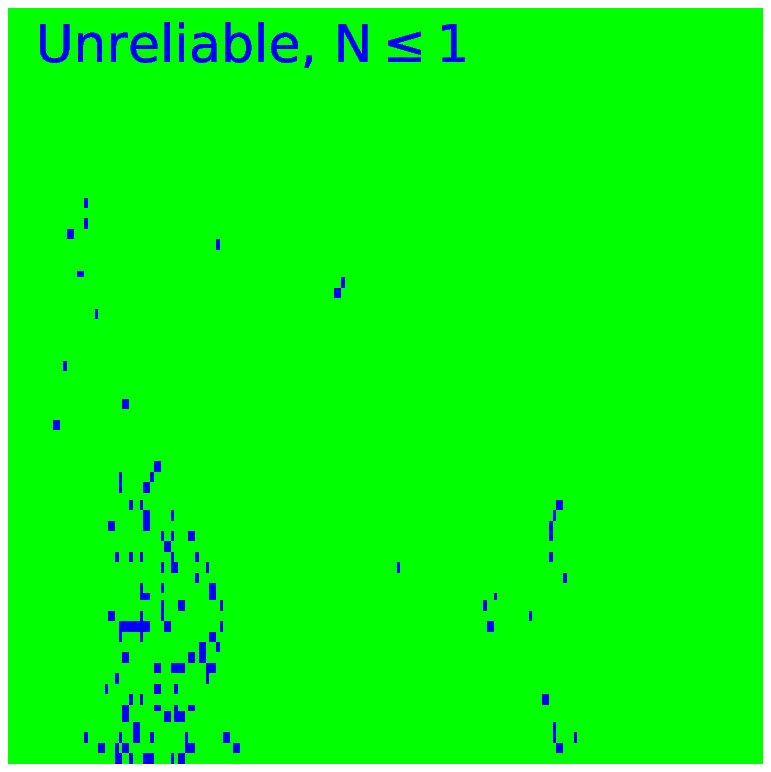}
\includegraphics[width=0.7\columnwidth]{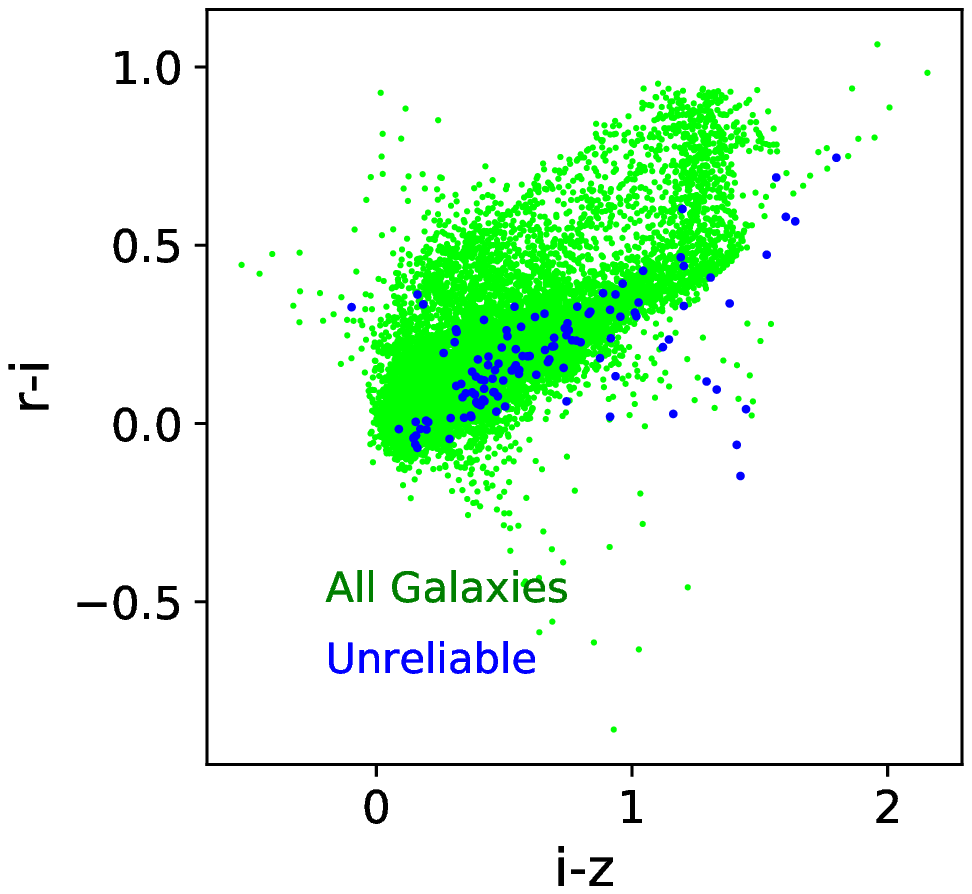}
\includegraphics[width=0.7\columnwidth]{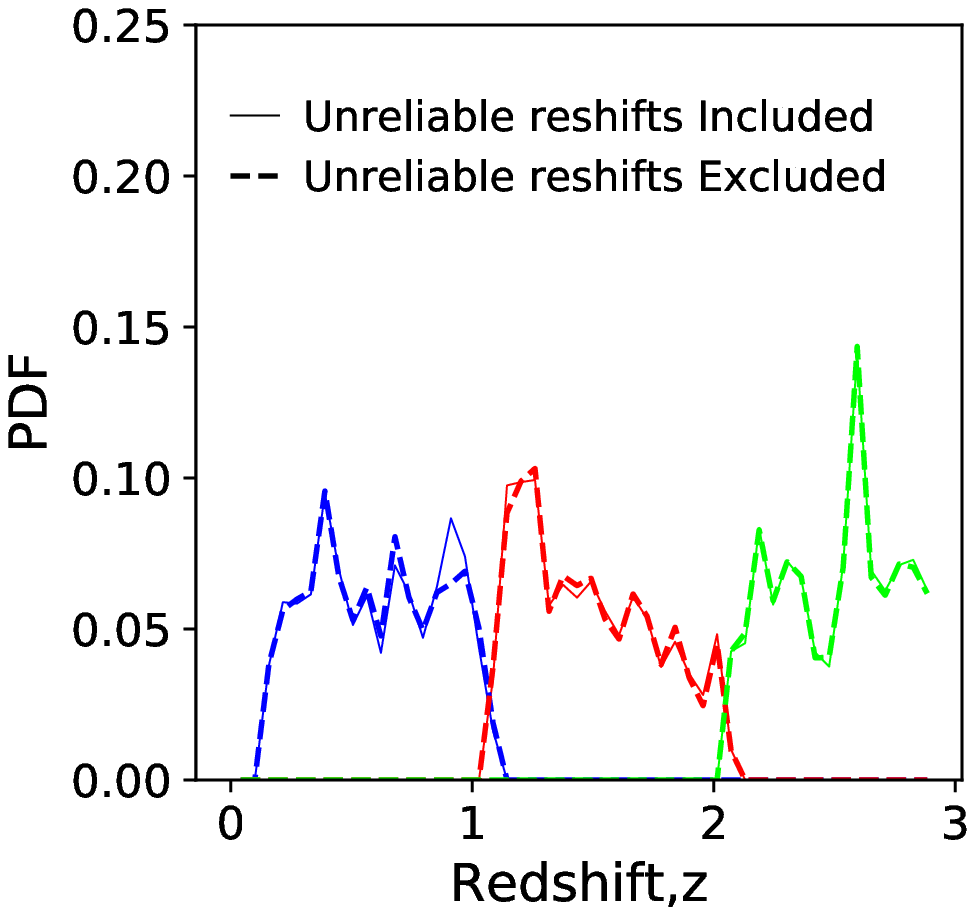}
\caption{Top: The self-organising map with pixels highlighted in blue that contain less than or equal to one galaxy. Middle: The distribution of the galaxies with unreliable redshift distributions (blue) in colour space, compared to all galaxies (green). Bottom: The equal-spaced tomographic bin PDFs when the unreliable pixels are either included (thin solid lines) or excluded (thick dashed lines).}
\label{foutlier}
\end{figure*}

In principle each pixel in the self-organising map (or equivalently voxel in colour space) could be used as a distinct tomographic bin. In this approach, as shown by \cite{taylor}, the total 3D shear-field information would be fully captured. The utility of binning the data is therefore only one of computational nature and regularisation. For the case that one uses a small number of tomographic bins $N_{\rm tomo}\ll N_{\rm pixel}$, which is usually the case, there will a dramatic speed up of the computation of cosmic shear power spectra, that scale as $\mathcal{O}(N^2_{\rm tomo})$. One may also  expect that the behaviour of the statistics may be more Gaussian (due to the central limit theorem), and the use of lower dimensional binned data may improve the properties of correspondingly lower dimensional matrices in an analysis. Furthermore covariance matrices computed from simulated data will require fewer simulations \cite{covt} (the number of simulations scales as $N_{\rm tomo}^2$). None of these has been shown to cause intractable problems in the regime that $N_{\rm tomo}\simeq N_{\rm pixel}$, but for the purposes of this paper we will assume that tomographic bin labelling is a desired approximation of the data.  

\subsection{Treatment of Outliers and Unreliable Redshifts}

Compared to traditional approaches based on splitting the galaxy sample by photometric redshift, we can distinguish problematic voxels and exclude these from the analysis. These can take a variety of forms. Firstly where the corresponding $n_\alpha(z)$ is multi-modal, secondly voxels for which there is insufficient or missing spectroscopic information to calibrate the colour-redshift relation, and thirdly where $n_{\alpha}(z)$ differs significantly from that of neighbouring voxels. 

The first category, with multimodal $n_{\alpha}(z)$, known as `outliers', result in redshift overlap between tomographic bins, thus reducing the efficacy. The second, with insufficient data, correspond to `unreliable' voxels. The third correspond to the sharp transitions in Fig.~\ref{fig:som_z}, where even though the voxels may have unimodal redshift distributions, uncertainties in the photometry may lead to multi-modal distributions

None of these are ideal, but especially voxels for which the redshift calibration is uncertain should be excluded. This is difficult to achieve when dividing the sample based on simple colour selections that may have a non-trivial distribution in the $N_B!/2$-dimensional colour space. Individual colour voxels, however, could in principle be identified and simply excluded. The reason to use the self-organising map space is the same reason why this method was proposed in \cite{masters}, namely that the dimensionality of the problem is reduced from $N_B!/2$ to $2$, thereby making calculations and selections in this space tractable.

%In this reconstituting of a tomographic bin definition it also becomes clear that one can redefine what a redshift `outlier' is: a pixel/voxel for which there is insufficient or missing spectroscopic information to calibrate the colour-redshift relation in that cell.  However in such an analysis even uncalibrated cells could still be included in such an analysis; only the distributions associated with those cells would be very broad. This is more straightforward than making colour selections that may have a non-trivial distribution in the $N_B!/2$-dimensional colour space. Individual colour voxels, however, could in principle be identified and removed in a similar fashion. The reason to use the self-organising map space is the same reason why this method was proposed in \cite{masters}, namely that the dimensionality of the problem is reduced from $N_B!/2$ to $2$, thereby making calculations and selections in this space tractable.

To demonstrate the usefulness of rainbow tomography, we consider the identification of unreliable voxels, but we note that in general all three complications should be characterised for each voxel. To do so, we identify all self-organising map pixels with less than or equal to one galaxy and label these as \henk{unreliable} as an example; this is a plausible selection to make for this simple demonstration. In Figure \ref{foutlier} we show the \henk{unreliable} pixels in the self-organising map plane, where each pixel can be identified and excluded, and in an example colour space projection using the same data ($r-i$ vs. $i-z$), where the distribution is much more complex and overlaps in projection with the sample of pixels for which adequate redshift information is available. We also show the tomographic bin PDFs for an equal-spacing configuration for three bins, with and without the \henk{unreliable pixels} excluded. Even in this simple case the difference caused by the exclusion of these pixels causes a shift in the mean redshift of the bins of $\Delta z=0.004, 0.001, 0.003$ for the three bins respectively from lowest to highest redshift; which would be outside the requirement of $\Delta z\leq 0.002$ for a Stage-IV cosmic shear experiment  \cite{tksys}. 

This provides an example of the type of error that could occur in a standard treatment of outliers, where such an exclusion would be difficult (involving high-dimensional exclusion boundaries in colour space), compared to the self-organising map case in which the exclusion of outliers would be straightforward.
\begin{figure}
\centering
\includegraphics[width=1\columnwidth]{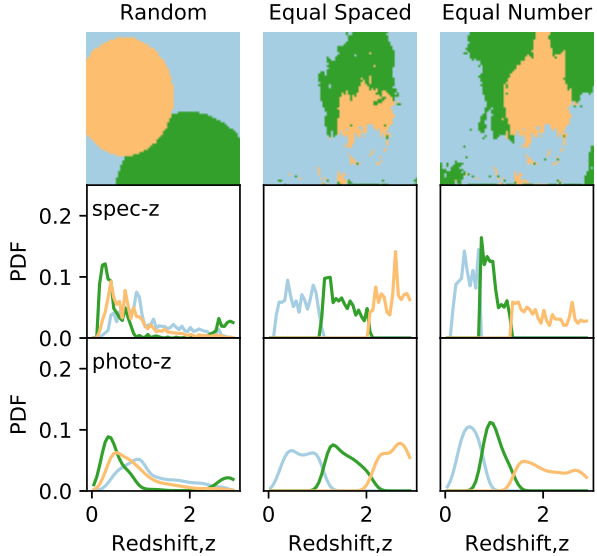}\\
\caption{The self-organising map with areas labelled by tomographic bin number for a 3-bin case where the examples of equally-spaced bins (right column), and equal-number bins (middle column) are shown. The left column is a random realisation of the label-space as defined in Section \ref{Optimisation Approach}. The upper panels show the self-organising maps where each colour labels a tomographic bin, the \red{middle} panels show the \red{spectroscopic redshift} probability density functions (PDFs) of the associated weights \red{$W_S\left(\eta_i, z \right)$} as a function of redshift, where each colour represents a different bin. \red{Similarly the lower panels show the photometric redshift probability density functions (PDFs) of the associated weights $W\left(\eta_i, z \right)$ as a function of redshift}. The colours match between panels.}
\label{fig:som_nz}
\end{figure}

\subsection{Optimisation Approach}
\label{Optimisation Approach}
%metric 
%only cosmic shear 
%searching algorithms
In Figure \ref{fig:som_nz} we show the same self-organising map as in Figure \ref{fig:som_z} with tomographic areas defined for the two cases of equally-spaced and equal-number bins. We show an illustrative case of 3 tomographic bins. It can be seen that the areas associated with each bin combination are very different, but they have a similar geometry. \red{We show the spectroscopic and photometric distributions for each bin choice, where it can be seen that the former is approximately a convolution of the latter; see equations (\ref{weight}) and (\ref{sweight}).} 

\begin{figure*}
\includegraphics[width=0.87\columnwidth]{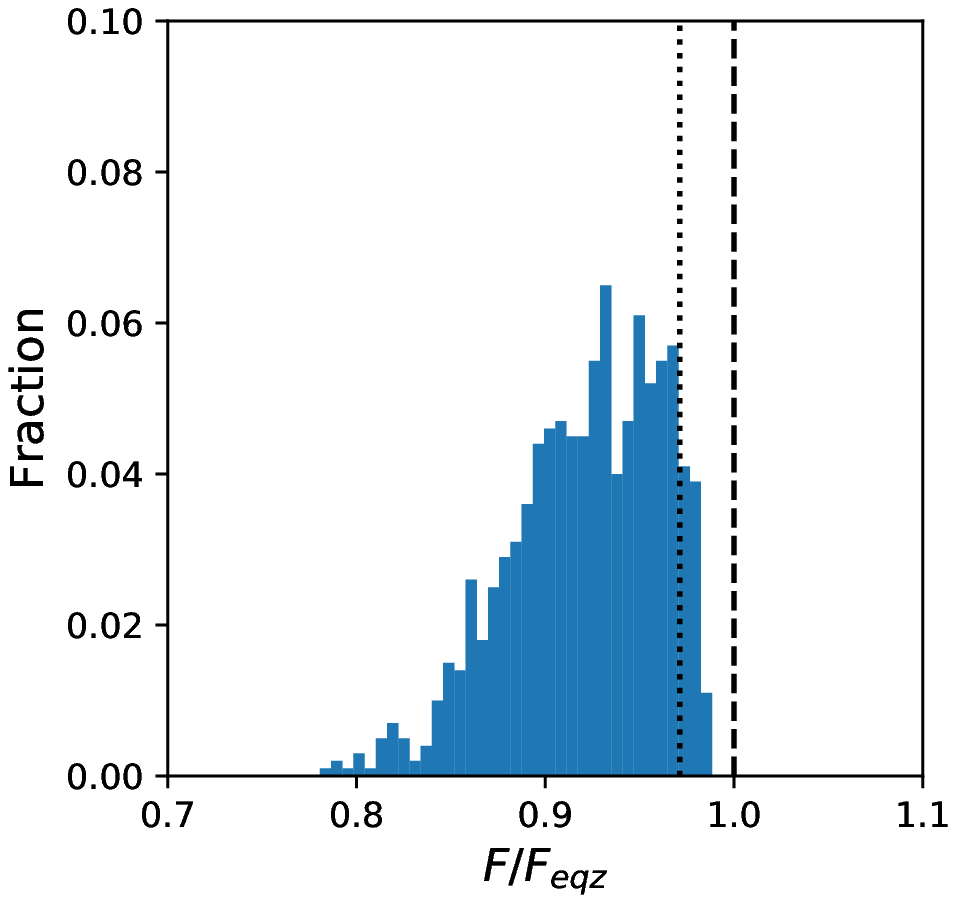}
\includegraphics[width=0.92\columnwidth]{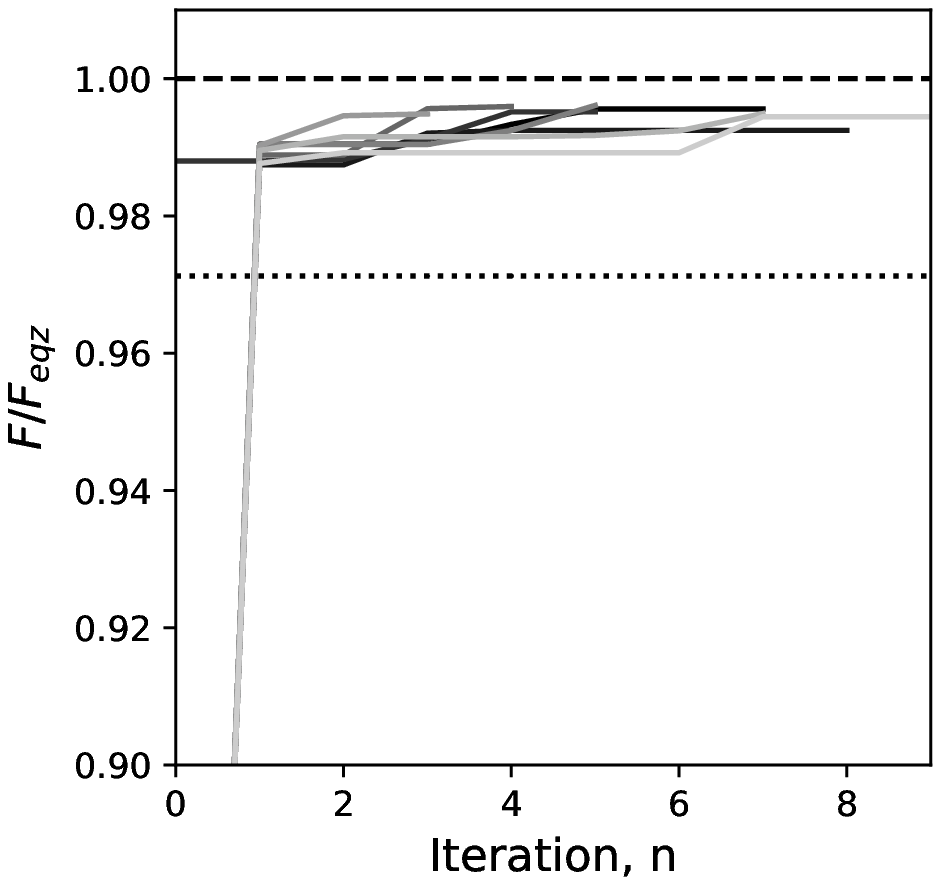}
\caption{The left panel shows the distribution of the optimisation metric, equation (\ref{eq:opt}), for 3 tomographic bin labels, when randomly sampling the label space. The dashed line is at $F/F_{\rm eqz}=1$, the equally-spaced cased, and the dotted line corresponds to the case of using equal-number binning. The right panel shows the convergence of the optimisation algorithm as a function of iteration number for 3 tomographic bin labels, again the dashed line is at $F/F_{\rm eqz}=1$ and the dotted line corresponds to the case of using an equal-number binning. The grey lines show different optimisation runs.}
\label{fig:som_opt3}
\end{figure*}
\begin{figure}
\includegraphics[width=1\columnwidth,clip=]{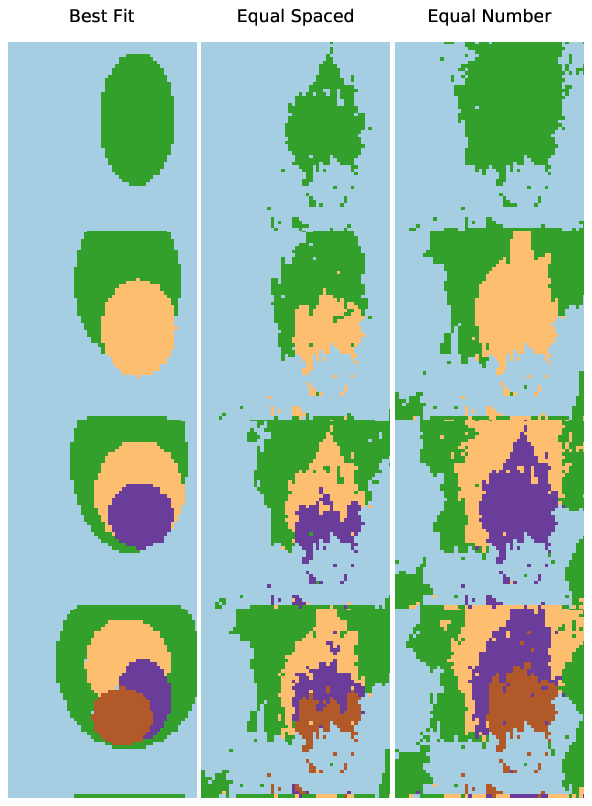}\\
\caption{The self-organising map with areas labelled by tomographic bin for 2,3,4 and 5-bin cases (top to bottom) where the left-hand panels show the best-fit solution using the label-space and optimisation routine described in Section \ref{Optimisation Approach}, the middle panels for equally-spaced bins, and the right-hand panels for equal-number bins. We suppress axes labels for clarity; as these are in arbitrary units. Each colour labels a different bin that can be identified in Figure \ref{fig:som_nz_opt_nz}.}
\label{fig:som_nz_opt}
\end{figure}

A full optimisation of the self-organising map for tomographic bin labelling would involve assigning a label between $[0,N_{\rm tomo}-1]$ for each self-organising map pixel and sampling all possible combinations of this labelling over all pixels. For $n\times m$ pixels this represents $(nm)^{N_{\rm tomo}-1}$ combinations. For the self-organising map we are using (dimension $(75, 150)$) for $N_{\rm tomo}>3$ tomographic labels the full brute-force searching of this space would have dimension $>1.4\times 10^{12}$. Even if one down-sampled this space by a factor $10$ in each dimension, thereby loosing sensitivity, this would still result in a space with $>1.4\times 10^6$ dimensions. Therefore this a brute-force optimisation is not possible. 

%describe the area definitions 
To make the bin labelling problem tractable, and to optimise over this space for $N_{\rm tomo}$ tomographic labels, we define $N_{\rm tomo}-1$ overlapping areas represented by elliptical disks. For each disk there are $4$ free hyper-parameters $(x,y,a,b)$ where $(x,y)$ are the centre of the disk in the self-organising map pixels, and $(a,b)$ are the semi-major and semi-minor axes of the disks (in principle this could be extended to include inclined ellipses with an orientation angle, but we found that this additional freedom was not required). For $N_{\rm tomo}$ tomographic labels this gives a `label-space' with dimensional $4(N_{\rm tomo}-1)$ (two labelled areas, and the unlabelled area, that defines the $N_{\rm tomo}^{\rm th}$ bin). Different combinations of the hyper-parameters describe different divisions of the self-organising map into tomographically binned labels. This is an extremely flexible space over which to optimise and allows one to optimise tomographic bin labelling in a way that is consistent with the underlying topography of the colour space (recall that neighbouring pixels have similar SEDs). In Figure \ref{fig:som_nz} we show an example of a random realisation of the hyper-parameter label space for three tomographic bin labels.   

%describe the metric
Given a particular labelling one needs to define a metric, or figure of merit, that quantifies how optimal this labelling is. To do this we use the cosmic shear signal-to-noise ratio for the dark energy parameter $w_0$ defined as
\begin{equation}
\label{eq:opt}
F=-\sum_{\ell,i,j}\left[\frac{\partial C_{\ell}(\eta_i,\eta_j)}{\partial w_0}\right]^2\frac{1}{[C_{\ell}(\eta_i,\eta_j)+N_{\ell}(\eta_i,\eta_j)]^2}, 
\end{equation}
where $C_{\ell}$ and $N_{\ell}$ are defined in equations (\ref{eq:c_l}) and (\ref{eq:Noise}) respectively. Throughout we present a normalised value of this quantity $F/F_{\rm eqz}$, where the denominator $F_{\rm eqz}$ is equal to $-F$ (note the minus sign to maintain the negativity of the optimisation metric, that is required for the optimisation algorithms we use) except using the equally-spaced bins (see e.g. Figure \ref{fig:som_nz}). $F$ is effectively (minus) the square root of the normalised $w_0$ component of the cosmic shear Fisher matrix. To compute the power spectra we use a Planck \cite{planck18} maximum likelihood cosmology. We note that many metrics could be used in this optimisation, and we leave an exploration of possible metric choices to future work. 

%describe the optimisation algorithm used 
The label space defined above is still relatively large, for $(3,4,5)$ tomographic labels the space has dimensions $(8,12,16)$, and due to the overlapping nature of the elliptical regions and sharp boundaries the space is also highly structured with many local extrema. Therefore an optimisation algorithm needs to be chosen that can cope with these conditions. To do this we chose the {\tt SciPy} differential evolution optimisation\footnote{\url{https://docs.scipy.org/doc/scipy-0.17.0/reference/generated/scipy.optimize.differential_evolution.html}} \cite{Storn1997}, which is a genetic algorithm that supports a population of points that mutate and evolve to the best-fit solution. Genetic algorithms are particularly suitable for highly structured optimisation problems with many local extrema. 

We also confirmed our results by using two other algorithms: the {\tt SciKit} optimisation package\footnote{\url{https://scikit-optimize.github.io}}, which uses either random forests, gradient descent boosting, or Gaussian mixture models to represent the optimisation surface more efficiently; and {\tt PyMultiNest}\footnote{\url{https://johannesbuchner.github.io/PyMultiNest/}} (\cite{pymultinest}), which uses nested sampling to search the optimisation space. We found for this particular optimisation problem that these approaches were slower; this is because we are only concerned with the best-fit solution and so algorithms that compute other quantities such as Bayesian evidence explore parts of the label space that are not necessary for our purposes. The code for this paper is available on request, where we use {\tt GLaSS} \cite{taylorflat} and {\tt CosmoSIS} \cite{cosmosis} to compute the cosmic shear power spectra. 

\section{Results} 
\label{sec:results}
% Equal-metric weights as an example 
% tomobins 2-5 optimal 
% example from bin 3 of convergence 
% ratio as a function of bin number 
In Figure~\ref{fig:som_opt3} we show random realisations of the label space for three tomographic bin labels if we randomly sample the (8 dimensional in this case) space. The left panel shows the distribution of $F/F_{\rm eqz}$. \red{The conclusions from this are two-fold. First} we find that there is a small spread in the distribution of the metric where the extreme values are between $[0.7,1.0]$; we confirmed this is the case for all $N_{\rm tomo}\leq 5$. This demonstrates that there is at most $30\%$ impact on dark energy sensitivity over a wide range of tomographic bin configurations. \red{Secondly this shows that over a wide space of possible bin configurations equally-spaced bins always perform better than an arbitrary bin configuration.} We find that the performance of equal-number binning is $\simeq 3\%$ lower than than the equally-spaced configuration. 
The right panel of Figure~ \ref{fig:som_opt3} shows the convergence of the optimisation algorithm as a function of iteration number when the {\tt SciPy} optimisation routine iterates to a global solution, over different realisations/sampling of the parameter space. We find that the algorithm always converges in fewer than $10$ iterations. 

\begin{figure*}
\centering
\qquad\;\;{\bf spec-z}\qquad\qquad\qquad\qquad\qquad\qquad\qquad\qquad\qquad\qquad\qquad\;\;\;\,\,{\bf photo-z}\\
\includegraphics[width=1\columnwidth]{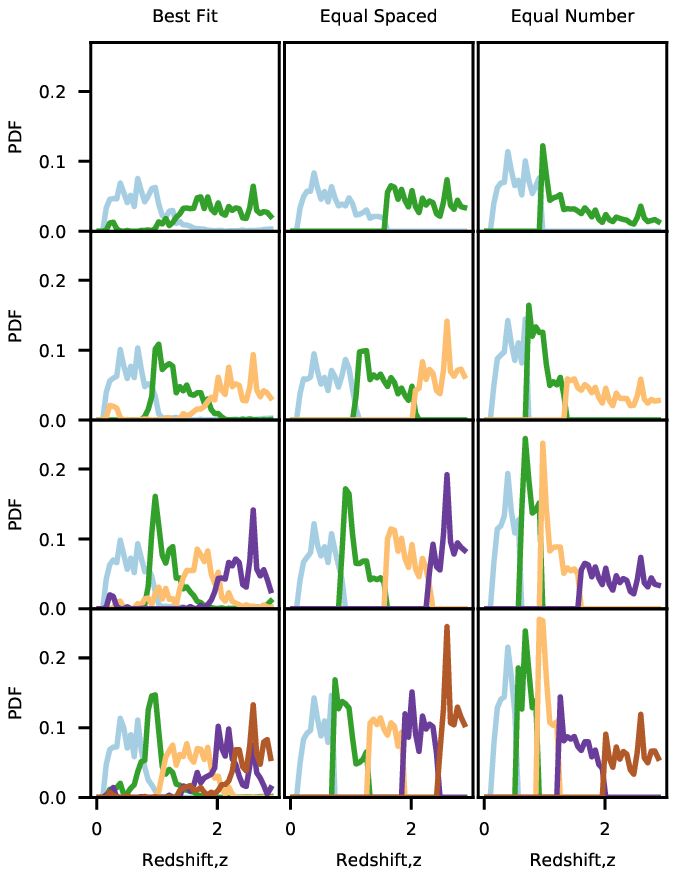}
\includegraphics[width=1\columnwidth]{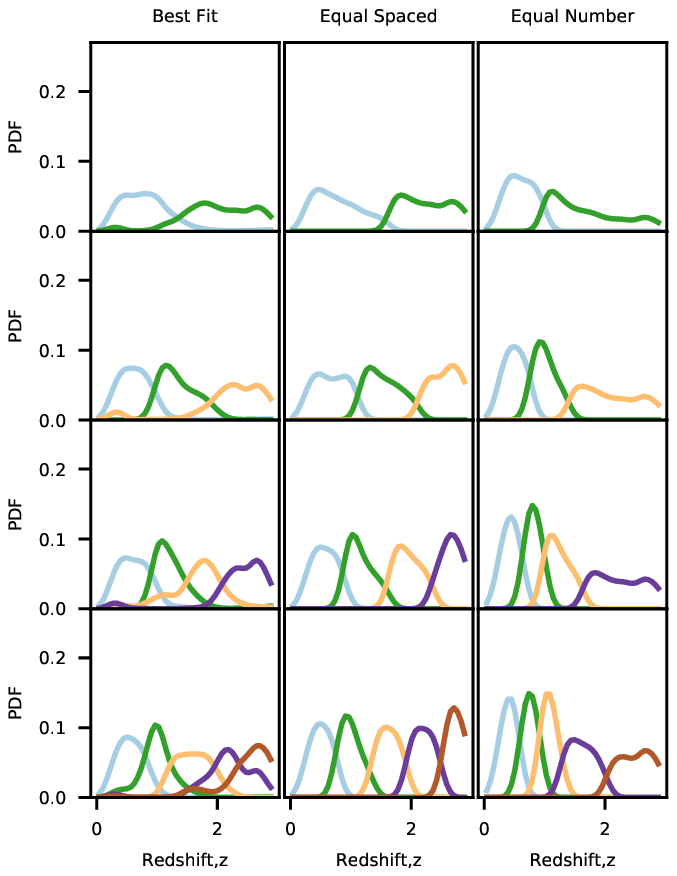}\\
\caption{The probability density functions (PDFs) of the associated weights $W\left(\eta_i, z \right)$ as a function of redshift corresponding to the labelling schemes in Figure \ref{fig:som_nz_opt}, where each colour represents a different bin. \red{The left hand panels show the spectroscopic redshift distributions and the right hand panels the associated photometric redshift distributions.} \red{In each plot} the left-hand panels show the best-fit solution using the label-space and optimisation routine described in Section \ref{Optimisation Approach}, the middle panels for equally-spaced bins, and the right-hand panels for equal-number bins. We suppress axes labels to allow more space; the y-axes are PDFs and the x-axes are redshift.}
\label{fig:som_nz_opt_nz}
\end{figure*}

In Figures \ref{fig:som_nz_opt} and \ref{fig:som_nz_opt_nz} we show the result of the optimisation for $N_{\rm tomo}=[2,5]$; for $N_{\rm tomo}>5$ all the optimisation algorithms become inefficient as they become prohibitively slow (as the dimension of the label space becomes large), and the ellipsoidal disk approximation becomes a poorer representation of the redshift boundaries in this space (see Figure \ref{fig:som_z}). We find that in all cases the best-fit solution closely matches the equally-spaced solution.

\begin{figure}
\centering
\includegraphics[width=0.9\columnwidth]{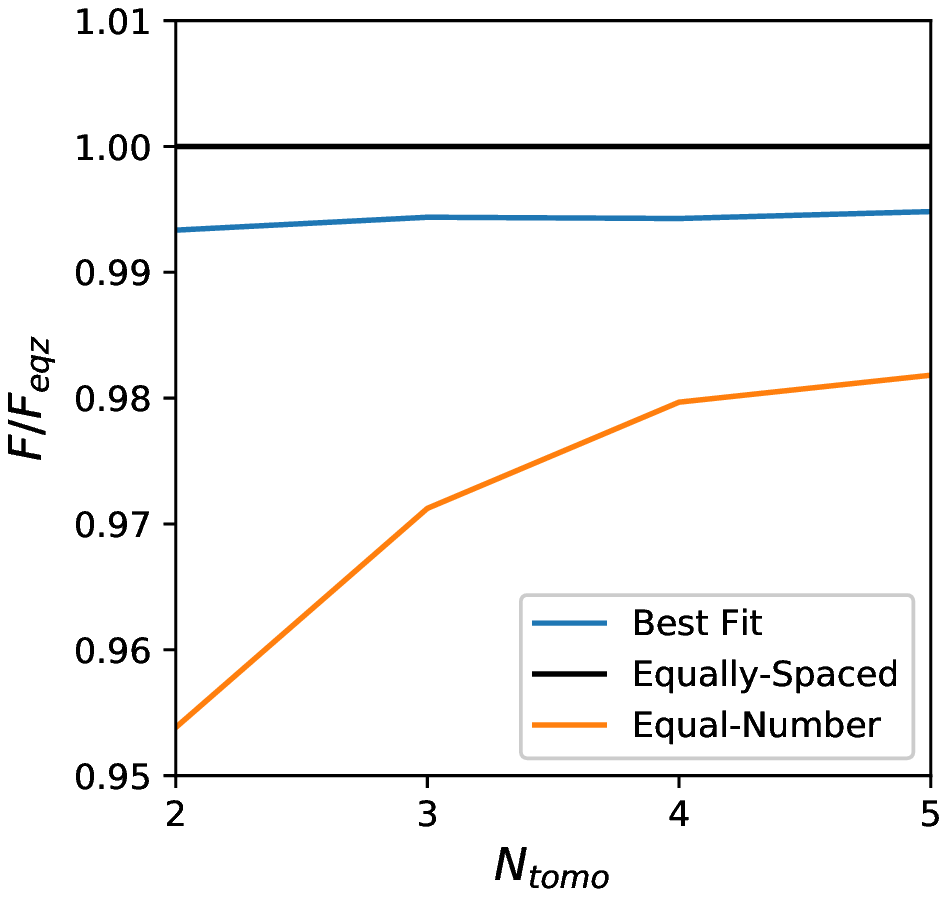}
\caption{The optimisation metric given in equation (\ref{eq:opt}) as a function of tomographic bin number. The lines show the equally-spaced in redshift case (black line) that is constant and equal to unity by definition; the best-fit solutions shown in Figure \ref{fig:som_nz_opt} and \ref{fig:som_nz_opt_nz} (blue line); and the equal-number density labelling scheme (orange line).}
\label{fig:opt}
\end{figure}

In Figure \ref{fig:opt} we show the metric as a function of $N_{\rm tomo}$. We find that the best-fit solution metric is relatively constant when the number of bins is increased, and that the equal-number labelling scheme is consistently lower with a slow convergence towards the equally-spaced binning. In all cases, as is also suggested by the random sampling shown in Figure \ref{fig:som_opt3}, the equally-spaced binning has a higher metric, suggesting that this is the global optimal solution. 

\subsection{Discussion}
% why equal-z is optimal - refer to Peter paper 
In this paper we investigate optimised tomographic bin labelling schemes by searching a label space that captures a wide variety of tomographic bin configurations. The optimisation metric we use is the dark energy equation of state signal-to-noise ratio for the cosmic shear power spectrum. In all cases we find a best-fit solution that is an approximation to a labelling scheme that has equally-spaced bins in redshift. Moreover, we find that the equally-spaced solution has a signal-to-noise ratio that is slightly higher than our best-fit solution. This is likely to be due to the approximate way in which we define areas in the self-organising map (using ellipsoidal disk features), which cannot accurately capture small disconnected regions with the same redshift. 

All of these conclusions suggest that defining equally-spaced bins in redshift \emph{is} the optimal configuration for a cosmic shear power spectrum analysis to maximise the dark energy equation of state signal-to-noise ratio. This is complementary to the conclusions of \cite{taylor} who find that for a sufficiently large number of bins both an equal-number and equal-spaced configuration captures all the 3D information from the shear field, but that as the number of bins increased the equal-number case converges more slowly than the equal-spaced case. 

The reason why equally-spaced bins are optimal is two-fold: the bins are orthogonal, and the equal-spacing provides better redshift coverage for dark energy measurements. Orthogonal bins should have a higher overall signal-to-noise ratio than non-orthogonal bins. If bins are non-orthogonal the overlap causes the cross-correlations between bins to have a non-zero noise component that is not present in the orthogonal case. Furthermore the signal is diluted in the redshift direction. For example, two completely overlapping bins would be probing exactly the same large-scale structure. The optimality of orthogonal bins is a result of the positivity of the weight functions we define in equation (\ref{weight}). 

\begin{figure}
\centering
\includegraphics[width=0.9\columnwidth]{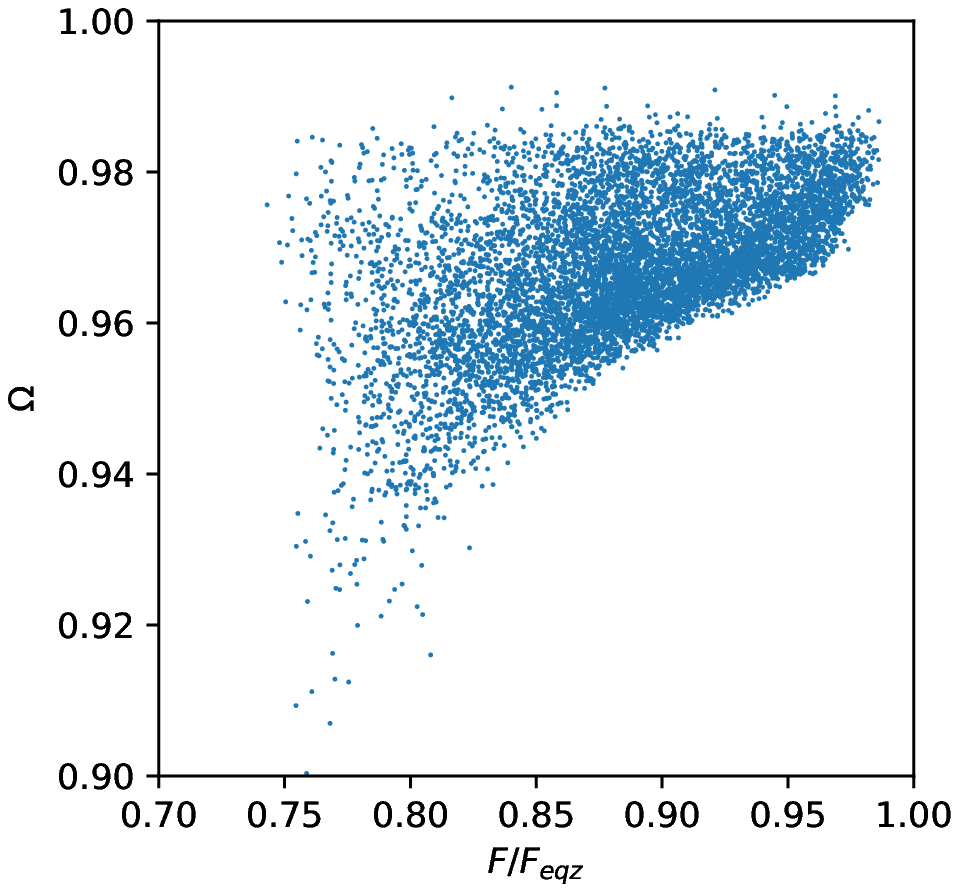}
\caption{The optimisation metric given in equation (\ref{eq:opt}) as a function of the orthogonality between the bins, defined in equation (\ref{orth}) for $10$,$000$ realisations of the tomographic bin configurations for $N_{\rm tomo}=3.$}
\label{fig:orth}
\end{figure}

To test the relationship between the orthogonality of the tomographic bins and the signal-to-noise ratio we define a measure of orthogonality as
\begin{equation}
\label{orth}
    \Omega=1-\frac{2}{N_{\rm tomo}(N_{\rm tomo}-1)}\sum_{\forall i\not=j} \frac{2W(\eta_i,z)W(\eta_j,z)}{W^2(\eta_i,z)+W^2(\eta_j,z)}
\end{equation}
where the sum is over all tomographic bin configurations. In this case $\Omega=1$ if all bins are orthogonal and $\Omega=0$ if they are completely overlapping. In Figure \ref{fig:orth} we show $\Omega$ as a function of the optimisation metric $F/F_{\rm eqz}$ for $10$,$000$ realisations of the tomographic bins configurations for $N_{\rm tomo}=3$. We find that indeed there is a strong relation between the orthogonality of the bins and the metric. For a given orthogonality there is a maximum metric that can be achieved, but also a distribution of configurations that are less optimal (consider the case of equally-space and equal-number, both are orthogonal but one is less optimal).

We make a number of approximations in this study that can be relaxed in future work. We focus here on optimisation of cosmic shear power spectra as a proof of concept, but this should be extended to include a larger number of tomographic bins, intrinsic alignment power spectra and a joint optimisation with galaxy-clustering methods. Furthermore the metric could be extended to include a full Fisher matrix for a cosmological parameter set rather than a single cosmological parameter.
\\

\section{Conclusion}
\label{Conclusion}
In this paper we generalise cosmic shear tomography, in which tomographic bins are traditionally defined directly in redshift space, to the case where tomographic bins are defined in photometric colour space where each colour space voxel is labelled by a bin number; this is similar to that proposed in \cite{jain}. In this case the redshift distribution of the bins is the sum of the redshift probability distributions of the galaxy populations given the same tomographic bin label. This results in weight functions that can overlap in redshift in complex ways. We demonstrate that in this redefinition it is more straightforward to exclude \henk{galaxies with undesirable redshift distributions} from a photometric redshift sample. 

We then define an approach to find optimal tomographic bin labelling. To do this we use the self-organising map of \cite{masters} that already compresses the $N_B!/2$-dimensional colour space for $N_B$ broad photometric bands down into a $2$ dimensional space. We then define regions in this space represented by ellipsoidal areas that for $N_{\rm tomo}$ tomographic bins creates a `label space' of dimension $4(N_{\rm tomo}-1)$. By searching over this label space we can optimise tomographic bin labelling for any metric that depends on the tomographic weight functions. 
\\

We define the dark energy equation of state signal-to-noise ratio for the cosmic shear power spectrum as our metric. To perform the optimisation we use the {\tt SciPy} differential evolution algorithm \cite{Storn1997}. We find that there is at most a 30\% sensitivity in the dark energy equation of state signal-to-noise ratio to the tomographic bin configuration; this is in agreement with \cite{hoyle} who find cosmological results to be largely insensitive to tomographic bin configuration. We find that for $N_{\rm tomo}\leq 5$ in all cases the best-fit solution is a close approximation to the case where equally-spaced bins in redshift are defined. Moreover, the equally-spaced bin configuration outperforms the best-fit solution given the label space we use. This suggests that defining equally-spaced bins in redshift is the optimal binning strategy for the metric we use. 

This study can be extended to include further statistics such as photometric galaxy clustering measurements, as well as systematic effects such as intrinsic alignments. By generalising the definition of cosmic shear tomography to be labels in colour-space, rather than bins in redshift, this opens up the possibility of optimising tomographic bin configurations to maximise the science return from future experiments.
\\

\section*{Acknowledgements}
We thank the {\tt Cosmosis} team for making their code publicly available. P.T. is supported by STFC. T.K. is supported by a Royal Society University Research Fellowship. We thank J. Zuntz for providing a development branch of {\tt Cosmosis} that was used in this analysis. D.M. and P.C. acknowledge support by NASA ROSES grant 12-EUCLID12-0004. D.M. acknowledges support for this work from a NASA Postdoctoral Program Fellowship. H.H. acknowledges support from Vici grant 639.043.512, financed by the Netherlands Organisation for Scientific Research (NWO).
\\
\newpage
\bibliographystyle{apsrev4-1.bst}
\bibliography{main.bib}

\end{document}